\documentclass[twocolumn]{revtex4}

\usepackage{graphicx}
\usepackage{dcolumn}

\newcommand{\beq}{\begin{eqnarray}}
\newcommand{\eeq}{\end{eqnarray}}

\renewcommand{\d}{\partial}

\def\Ren#1{\mbox{\boldmath$#1$}}

\begin{document}

\preprint{}

\title{
Uniqueness of Landau-Lifshitz Energy Frame in Relativistic Dissipative Hydrodynamics
}

\author{Kyosuke Tsumura}
\affiliation{
Analysis Technology Center,
Fujifilm Corporation,
Kanagawa 250-0193, Japan
}

\author{Teiji Kunihiro}
\affiliation{
Department of Physics,
Kyoto University,
Kyoto 606-8502, Japan
}

\date{\today}

\begin{abstract}
We show that
the relativistic dissipative hydrodynamic equation
derived from the relativistic 
Boltzmann equation by the renormalization-group method
uniquely leads to the one in the energy frame proposed by Landau and Lifshitz,
provided that the macroscopic-frame vector, which
defines the local rest frame of the fluid velocity,
is independent of the momenta of constituent particles, as it should.
We argue that
the relativistic hydrodynamic equations for viscous fluids
must be defined on the energy frame if it is consistent with the underlying
relativistic kinetic equation.
\end{abstract}

\maketitle

\setcounter{equation}{0}
\section{
   Introduction
}
\label{sec:001}
Theory of relativistic hydrodynamics for viscous fluids
is a powerful 
mean
for analyzing the long-wavelength and low-frequency dynamics of many body systems,
such as
the hot and/or dense quark-gluon or hadronic matter
created by relativistic heavy-ion collisions \cite{qcd001,qcd003,Bozek:2011gq,Hirano:2012kj}
and also
various high-energy astrophysical phenomena \cite{ast001,ast002}.

However, 
the relativistic dissipative hydrodynamic equation 
is still under debate in the fundamental level.
Indeed, the following three problems may be noted:
(A) There are ambiguities in the definition of the fluid velocity \cite{hen001,hen002},
(B) In the Eckart (particle) frame, there arises an unphysical
 instabilities of the equilibrium state \cite{hyd002},
and
(C) the so-called first-order equations lack in causality \cite{mic001,hen003,mic004}.

As is easily thought of,
it is a legitimate and natural way 
for  obtaining the proper relativistic hydrodynamic equation, 
to start with 
the relativistic Boltzmann equation (RBE)
 which is Lorentz invariant and does not have  stability nor causality problems \cite{mic001}.
The problem is how to 
obtain an asymptotic  dynamics in the far-infrared long-wavelength limit
of the RBE, and several reduction methods \cite{chapman,grad,Kampen} have been developed with 
some but not a compete success.

Recently,
for examining the first two problems, (A) and (B),
Ohnishi and the present authors \cite{env009,Tsumura:2011cj}
applied the renormalization-group (RG) method \cite{rgm001,env001,env006,chiba}
as a powerful reduction theory of dynamics
to  the RBE
and succeeded in deriving a generic form of relativistic dissipative hydrodynamic equations
with the stable equilibrium state.
A key ingredient in the derivation was
introduction of a time-like Lorentz-covariant vector $\Ren{a}^\mu$, with $\Ren{a}^0 > 0$:
$\Ren{a}^\mu$ specifies the macroscopic and covariant coordinate system
where the local rest frame of
the fluid velocity $u^\mu$ is defined,
and is called the macroscopic-frame vector.
$\Ren{a}^\mu$ could depend on the momenta $p^\mu$ of constituent particles
of the system as well as the space-time coordinate $x^\mu$.
In fact,
Ohnishi and the present authors \cite{env009,Tsumura:2011cj}
 could manage to derive 
the hydrodynamic equation in the Eckart (particle) frame 
only by making $\Ren{a}^\mu$ have a $p^\mu$ dependence.
In retrospect, the possible momentum dependence of $\Ren{a}^\mu$, however, 
may not be legitimate for 
$\Ren{a}^\mu $ to play a macroscopic-frame vector,
because it means that the macroscopic space-time is defined 
for respective particle states with a definite energy-momentum,
and may lead to a difficulty in the physical interpretation of the space and time
in which the hydrodynamics is defined. 
Thus, we are lead to  require that 
$\Ren{a}^\mu$ should be independent of $p^\mu$ and time-like vector with the Lorentz covariance.

In this paper, taking this requirement from the outset, 
we shall examine the outcomes and reach a significant conclusion that
the relativistic hydrodynamics consistent with the
underlying kinetic equation should be uniquely the one 
in the Landau-Lifshitz (energy) frame,
but not in other frames.
This means also that 
the so-called matching conditions for selecting the energy frame but not 
other frames \cite{mic001}
is uniquely derived from the underlying kinetic equation \cite{EPJA}.

This paper is organized as follows:
After 
a brief account of the basic properties of
the relativistic Boltzmann equation (RBE),
we show that $\Ren{a}^\mu$ that is independent of the momentum $p^\mu$
must be naturally 
proportional to the fluid velocity $u^\mu$;
$\Ren{a}^\mu = b \, u^\mu$.
Furthermore,
we show that
the ``normalization'' factor $b$ can be made unity without loss of generality.
Then,
we apply the RG method to 
derive  the relativistic hydrodynamics 
from the RBE 
with
$\Ren{a}^\mu = u^\mu$, and 
 show that the resulting hydrodynamic equation is
uniquely in the energy frame. We also clarify the meaning of the matching condition
in terms of the inner product for the distribution functions.
The last section is devoted to a summary and concluding remarks
on the cases of multi-component systems and the extended thermodynamics.

\setcounter{equation}{0}
\section{
   Relativistic Boltzmann equation
}
\label{sec:002}
To make the presentation self-contained, we first summarize the basic facts about
the relativistic Boltzmann equation (RBE) \cite{mic001} very briefly.
The RBE is an evolution equation of the one-particle distribution function
$f_p(x)$ defined in the phase space $(x, \, p)$, 
\begin{eqnarray}
  \label{eq:1-001}
  p^\mu  \partial_\mu f_p(x) = C[f]_p(x),
\end{eqnarray}
with  $p^{\mu}$ being the four-momentum of the on-shell particle,
\textit{i.e.}, $p^\mu p_\mu = p^2 = m^2$ and $p^0 > 0$.
The $C[f]_p(x)$ in the right hand side denotes the collision operator,
\begin{eqnarray}
  \label{eq:1-002}
  C[f]_p(x) &\equiv& \frac{1}{2!} \, \sum_{p_1} \, \frac{1}{p_1^0} \,
  \sum_{p_2} \, \frac{1}{p_2^0} \, \sum_{p_3} \, \frac{1}{p_3^0} \,
  \omega(p, p_1|p_2, p_3)\nonumber\\
  &&{}\times\Big( f_{p_2}(x) f_{p_3}(x) - f_p(x) f_{p_1}(x) \Big),
\end{eqnarray}
with $\omega(p \,,\, p_1|p_2 \,,\, p_3)$ being the transition probability,
which has the symmetry property
$\omega(p, p_1|p_2, p_3) = \omega(p_2, p_3|p, p_1)
= \omega(p_1, p|p_3, p_2) = \omega(p_3 , p_2|p_1 , p)$
and respects 
the energy-momentum conservation
$\omega(p, p_1|p_2, p_3) \propto \delta^4(p + p_1 - p_2 - p_3)$.

Thanks to the above symmetry property and the energy-momentum conservation,
the function $\varphi_p(x)=a(x) + p^{\mu} \, b_{\mu}(x)$ 
is found to be a collision invariant;
\begin{eqnarray}
  \label{eq:1-005}
  \sum_p \, \frac{1}{p^0} \, \varphi_p(x) \, C[f]_p(x) = 0,
\end{eqnarray}
where $a(x)$ and $b_{\mu}(x)$ being arbitrary functions of $x$.
On account of Eq. (\ref{eq:1-005}),
we have formally the balance equations
for the energy-momentum tensor 
$T^{\mu\nu}(x)=\sum_p \, \frac{1}{p^0} \, p^\mu \, p^\nu \, f_p(x)$
and the particle current 
$N^\mu(x)=\sum_p \, \frac{1}{p^0} \, p^\mu \, f_p(x)$ as follows;
  $\partial_\nu T^{\mu\nu}(x)=0$
and
$\partial_\mu N^\mu(x)=0$, respectively. 
Despite the appearance,
 these equations do not have any dynamical information
before the evolution of $f_p(x)$
has been obtained from Eq. (\ref{eq:1-001}).

The entropy current is defined by
 $ S^\mu(x) \equiv - \sum_p  \frac{1}{p^0} p^\mu  f_p(x)
  (\ln (2\,\pi)^3 \, f_p(x) - 1)$,
which is generically non-conserving;
\begin{eqnarray}
  \label{eq:1-010}
  \partial_\mu S^\mu(x) =  - \sum_p \, \frac{1}{p^0} \, C[f]_p(x) \, \ln (2\,\pi)^3 \, f_p(x),
\end{eqnarray}
due to Eq.(\ref{eq:1-001}).
One sees that
$S^\mu$ becomes a conserved quantity only when 
$\ln (2\,\pi)^3 \, f_p(x)$ is a collision invariant,
$\ln (2\,\pi)^3 \, f_p(x) = a(x) + p^\mu \, b_\mu(x)$.
Accordingly, an entropy-conserving 
 distribution function can be cast into the form of
the local equilibrium distribution function (the J\"uttner function \cite{juettner})
in terms of the local temperature $T(x)$, chemical potential $\mu(x)$, and 
 time-like fluid velocity $u^\mu(x)$
as
 $ f_p(x) = (2\pi)^{-3}
  \exp [ ({\mu(x) - p^\mu  u_\mu(x)})/{T(x)}]  \equiv f^{\mathrm{eq}}_p(x)$,
with $u^\mu(x) \, u_\mu(x) = 1$ and $u^0(x) > 0$.
We also note that
the collision operator identically vanishes 
for the {\em local} equilibrium distribution function $f^{\mathrm{eq}}_p(x)$;
\begin{eqnarray}
  \label{eq:1-012}
  C[f^{\mathrm{eq}}]_p(x) = 0,
\end{eqnarray}
due to the energy-momentum conservation in the collision process.

\setcounter{equation}{0}
\section{
  Macroscopic-frame vector
}
\label{sec:003}

To implement the task to  solve the RBE in the hydrodynamic regime,
it is customary \cite{mic001,env009,Tsumura:2011cj}
to introduce a time-like Lorentz vector $\Ren{a}^\mu$, with $\Ren{a}^0 > 0$.
We call $\Ren{a}^\mu$ the macroscopic-frame vector,
which defines the covariant and macroscopic coordinate system $(\tau,\,\sigma^\mu)$
from the space-time coordinate $x^\mu$ as
\begin{eqnarray}
  \label{eq:coodinates1}
  d\tau &=& \Ren{a}^\mu \, dx_\mu,\\
  \label{eq:coodinates2}
  d\sigma^\mu &=& \Big( g^{\mu\nu} - \frac{\Ren{a}^\mu\,\Ren{a}^\nu}{\Ren{a}^2} \Big) \, dx_\nu.
\end{eqnarray}
It is noted that
$(\tau,\,\sigma^\mu)$
is the so-called local Lorentz frame
when $\Ren{a}^\mu$ is dependent on $x^\mu$.

We now show that $\Ren{a}^\mu $ must be proportional to the fluid velocity $u^{\mu}$,
provided that 
$\Ren{a}^\mu$ should be independent of the momentum $p^\mu$ and time-like vector
with the Lorentz covariance. 
Noting that $u^\mu$ and $\partial^\mu$ are the only available Lorentz vectors at hand,
we see that the generic form of a Lorentz-covariant vector reads
\begin{eqnarray}
  \label{eq:general-a}
  \Ren{a}^\mu = A_1\,u^\mu + A_2\,\partial^\mu T
  + A_3\,\partial^\mu \mu + A_4\,u^\nu\,\partial_\nu u^\mu,
\end{eqnarray}
where $A_i$ with $i=1,\,2,\,3,\,4$ is an arbitrary Lorentz-scalar function of 
the temperature and the chemical potential; $A_i = A_i(T,\,\mu)$.
Now the derivative $ \partial^\mu$ can be decomposed into the time-like and
space-like components as
 $ \partial^\mu = u^\mu \, u^\nu \, \partial_\nu
  +(g^{\mu\nu} - u^\mu \, u^\nu) \, \partial_\nu
  \equiv u^\mu \, D + \nabla^\mu$,
where $D \equiv u^{\nu} \, \partial_\nu$
and $\nabla^\mu \equiv \Delta^{\mu\nu} \, \partial_\nu$
with $\Delta^{\mu\nu} = (g^{\mu\nu} - u^\mu \, u^\nu)$
being the projection operator onto a space-like vector
orthogonal to the time-like vector $u^\mu$.
Then 
Eq.(\ref{eq:general-a}) is also decomposed as
\begin{eqnarray}
  \Ren{a}^\mu &=&
  (A_1 + A_2 \, DT + A_3 \, D\mu) \, u^\mu\nonumber\\
  &&{}+ A_2 \, \nabla^\mu T + A_3 \, \nabla^\mu\mu + A_4 \, Du^\mu.
\end{eqnarray}
But since $\Ren{a}^\mu$ should be time-like,  we have
\begin{eqnarray}
  \Ren{a}^\mu =
  (A_1 + A_2 \, DT + A_3 \, D\mu) \, u^\mu 
  \equiv b(T,\,\mu) \, u^\mu,
\end{eqnarray}
with $b(T,\,\mu) > 0$.
Here, we have used the fact that
$\nabla^\mu T$, $\nabla^\mu \mu$, and $D u^\mu$ are space-like
because
$\Delta^{\mu\nu}\,\nabla_\nu = \nabla^\mu$
and
$\Delta^{\mu\nu}\,Du_\nu = Du^\mu$.

With the use of this macroscopic-frame vector,
the covariant and macroscopic coordinate system $(\tau,\,\sigma^\mu)$
reads
 $ d\tau = b(T,\,\mu)\,u^\mu \, dx_\mu$ and 
 $d\sigma^\mu = \Delta^{\mu\nu} \, dx_\nu$.
In fact,
the ``normalization'' factor $b(T,\,\mu)$ is a redundant degree of freedom
for the dynamics because
the $b(T,\,\mu)$ can be always made unity by converting
$\tau$ into the new temporal coordinate $\tau^\prime$ as
 $ d\tau^\prime \equiv b(T,\,\mu)^{-1} \, d\tau = u^\mu \, dx_\mu$.
From now on,
we thus set 
\begin{eqnarray}
  \Ren{a}^\mu=u^{\mu}.
\end{eqnarray}

In terms of the new coordinates $(\tau,\,\sigma^\mu)$, Eq.(\ref{eq:1-001}) is rewritten  as
\begin{eqnarray}
  \label{eq:start}
  \frac{\partial}{\partial \tau} f_p(\tau,\, \sigma)
  &=& \frac{1}{p \cdot u} C[f]_p(\tau,\, \sigma)\nonumber\\
  &&{}- \varepsilon \, \frac{1}{p \cdot u}
  \, p \cdot \nabla f_p(\tau,\, \sigma),
\end{eqnarray}
where ${\partial}/{\partial\tau} \equiv D$ and
${\d}/{\d \sigma_{\mu}} \equiv \nabla^\mu$.
Here the small parameter $\varepsilon$, which will be set back to unity eventually,
represents the non-uniformity of space.
This seemingly mere rewrite of the equation reflects a physical assumption
that only the spatial inhomogeneity is the origin of the dissipation.
We note that the RG method applied to the non-relativistic case with this assumption  
 successfully leads 
to the Navier-Stokes equation \cite{env007},
which means that
the present approach \cite{env009,Tsumura:2011cj}
is a covariantization of the non-relativistic case.

\setcounter{equation}{0}
\section{
   Reduction to hydrodynamic equation
}
\label{sec:004}

In this section,
we  derive
 the relativistic hydrodynamics 
through 
solving the converted Boltzmann equation (\ref{eq:start})
in the RG method.

We should note that some of formulas presented in this section may be 
found in the previous papers \cite{env009,Tsumura:2011cj} in a different context and 
hence in a different order. 
Thus, the presentation will be a brief reorganization of them;
we refer to  Ref. \cite{Tsumura:2011cj} for the detailed reasoning.

\subsection{
  Hydrodynamics from relativistic Boltzmann equation by RG method
}

In the RG method developed in Ref.'s \cite{env001,env006,env009,Tsumura:2011cj},
we first try to obtain the perturbative solution $\tilde{f}_p$ 
to Eq. (\ref{eq:start})
around the arbitrary initial time $\tau = \tau_0$
with the initial value $f_p(\tau_0,\,\sigma)$;
 $ \tilde{f}_p(\tau = \tau_0,\,\sigma;\,\tau_0) = f_p(\tau_0,\,\sigma)$.
Note that the solution depends on the initial time $\tau_0$ at which
$\tilde{f}_p(\tau = \tau_0,\,\sigma;\,\tau_0)$
is supposed to be on an exact solution.
We expand the initial value as well as the solution
with respect to $\varepsilon$ as follows:
\begin{eqnarray}
  \tilde{f}_p(\tau,\,\sigma;\,\tau_0)
  &=& \sum_{i=0}^{\infty} \, \varepsilon^i \, \tilde{f}_p^{(i)}(\tau,\,\sigma;\,\tau_0), \\
  f_p(\tau_0,\,\sigma)
  &=& \sum_{i=0}^{\infty} \, \varepsilon^i \, f_p^{(i)}(\tau_0,\,\sigma).
\end{eqnarray}
    
The zeroth-order equation is
\begin{eqnarray}
  \label{eq:1-025}
  \frac{\partial}{\partial \tau} \tilde{f}^{(0)}_p(\tau,\,\sigma;\,\tau_0)
  = \frac{1}{p \cdot u} \,
  C[\tilde{f}^{(0)}]_p(\tau,\,\sigma;\,\tau_0).
\end{eqnarray}
Since we are interested in the slow motion
realized asymptotically for $\tau \rightarrow \infty$,
we take the stationary solution satisfying
${\partial}\tilde{f}_p^{(0)}(\tau,\,\sigma;\,\tau_0)/{\partial \tau} = 0$,
which demands  that
\beq
C[\tilde{f}^{(0)}]_p(\tau,\,\sigma;\,\tau_0) = 0, \quad \,\, \forall\sigma.
\eeq
Such a solution is given by
a local equilibrium distribution function,
\textit{i.e.}, the J\"uttner distribution function,
\begin{eqnarray}
  \label{eq:1-028}
  \tilde{f}_p^{(0)}(\tau,\,\sigma;\,\tau_0)
  &=& \frac{1}{(2\pi)^{3}} \,
  \exp \Bigg[ \frac{\mu(\sigma;\,\tau_0)
  - p^\mu \, u_\mu(\sigma;\,\tau_0)}{T(\sigma;\,\tau_0)} \Bigg]\nonumber\\
  &\equiv& f^{\mathrm{eq}}_p(\sigma;\,\tau_0),
\end{eqnarray}
with $u^\mu(\sigma;\,\tau_0) \, u_\mu(\sigma;\,\tau_0) = 1$.
We note that the integration constants
$T(\sigma;\,\tau_0)$, $\mu(\sigma;\,\tau_0)$, and $u_\mu(\sigma;\,\tau_0)$
are independent of $\tau$ but may depend on $\tau_0$ as well as $\sigma$.
  
The first-order equation is given by
\begin{eqnarray}
  \label{eq:1-030}
  \frac{\partial}{\partial \tau} \tilde{f}_p^{(1)}(\tau)
  = \sum_q \, A_{pq} \, \tilde{f}_q^{(1)}(\tau) + F_p,
\end{eqnarray}
where 
\begin{eqnarray}
  F_p\equiv -  \frac{1}{p \cdot u} \, p \cdot \nabla f^{\mathrm{eq}}_p.
\end{eqnarray}
with $A_{pq}$ being a matrix element of
 the linearized collision operator $A$,
\begin{eqnarray}
  \label{eq:1-031}
  (A)_{pq}=A_{pq} \equiv \frac{1}{p \cdot u} \,
  \frac{\partial}{\partial f_q} C[f]_p \, \Bigg|_{f =  f^{\mathrm{eq}}}.
\end{eqnarray}

We now show that the linearized collision operator $A$ has  remarkable spectral
properties \cite{env009,Tsumura:2011cj}, which are essential for the following analysis.
We first  convert $A$ to another linear operator,
\begin{eqnarray}
 L \equiv (f^{\mathrm{eq}})^{-1} \, A \, f^{\mathrm{eq}},
\end{eqnarray}
with 
$(f^\mathrm{eq})_{pq} \equiv f^\mathrm{eq}_p \, \delta_{pq}$,
which is diagonal.
Next,
an inner product is introduced 
for arbitrary nonzero vectors $\varphi$ and $\psi$ as
\begin{eqnarray}
  \label{eq:1-034}
  \langle  \, \varphi \,,\, \psi \, \rangle
  \equiv \sum_{p} \, \frac{1}{p^0} \, (p \cdot u) \,
  f^{\mathrm{eq}}_p \, \varphi_p \, \psi_p,
\end{eqnarray}
which satisfies the positive-definiteness of the norm as
 $ \langle  \varphi,\,\varphi  \rangle =
  \sum_{p} \frac{1}{p^0} \, (p \cdot u) \,  f^{\mathrm{eq}}_p \, (\varphi_p)^2 >  0$,
for $\varphi_p \ne 0$,
due to $(p\cdot u) > 0$
because of both $p^{\mu}$ and $u^{\mu}$ being time-like vectors.

Then,
as is shown in Ref.'s \cite{env009,Tsumura:2011cj},
the linearized collision operator
is semi-negative definite and has five zero modes given by
\begin{eqnarray}
  \label{eq:1-039}
  \varphi_{0p}^\alpha \equiv \left\{
  \begin{array}{ll}
    \displaystyle{p^\mu} & \displaystyle{\mathrm{for}\,\,\,\alpha = \mu=0 \sim 3,} \\[2mm]
    \displaystyle{1\times m}     & \displaystyle{\mathrm{for}\,\,\,\alpha = 4}.
  \end{array}
  \right.
\end{eqnarray}
The functional subspace spanned by the five zero modes is called the P$_0$ space and
the projection operator to it is denoted by $P_0$; 
 $ \big[ P_0 \, \psi \big]_p \equiv
  \varphi_{0p}^\alpha \, \eta^{-1}_{0\alpha\beta} \,
  \langle \, \varphi_0^\beta \,,\, \psi \, \rangle$,
where
$\eta^{-1}_{0\alpha\beta}$ is the inverse matrix of
the the P${}_0$-space metric matrix $\eta_0^{\alpha\beta}$
defined by
 $ \eta_0^{\alpha\beta} \equiv \langle \, \varphi_0^\alpha \,,\, \varphi_0^\beta \, \rangle$.
We also call the complement to the P$_0$ space the Q$_0$ space and introduce $Q_0 \equiv 1 - P_0$.
In the following, we also use the modified projection operators 
defined by
 $ \bar{P}_0=f^{\rm eq}P_0(f^{\rm eq})^{-1}$ and
$  \bar{Q}_0=f^{\rm eq}Q_0(f^{\rm eq})^{-1}$.

Then,
the perturbative solution up to the second order reads
 $ \tilde{f}_p(\tau,\sigma;\tau_0)
  = \tilde{f}^{(0)}_p + \varepsilon \tilde{f}^{(1)}_p+ 
\varepsilon^2  \tilde{f}^{(2)}_p  + O(\varepsilon^3)$,
where
\beq
\tilde{f}^{(1)}_p
= (\tau - \tau_0) \bar{P}_0  F - A^{-1}  \bar{Q}_0  F,
\eeq
and we refer to Ref. \cite{Tsumura:2011cj} for a lengthy formula of
$\tilde{f}^{(2)}_p$, for the sake of space.
We note that secular terms are present in $\tilde{f}^{(i)}_p$ ($i=1,\,2,\,\cdots$) ,
which is caused by the zero modes of the linearized collision operator $A$.

Remarks are in order here:
In  the usual approaches \cite{mic001,Tsumura:2011cj} to derive hydrodynamic equations from 
RBE, so-called matching conditions  corresponding to a desired frame are imposed 
on the higher-order terms from the outset,
and the hydrodynamic equation in the desired frame including Eckart one is 
formally obtained. It is noteworthy that 
the matching conditions are also a condition to forbid 
 the appearance of secular terms \cite{mic001};
we refer to Ref. \cite{Tsumura:2011cj}
for some problematic aspects of the matching conditions.
In the present approach, on the other hand, 
we have so far just solved the Boltzmann equation 
in the perturbation theory in a straightforward way without imposing any matching conditions,
and the resulting higher-order terms contain secular terms, 
which apparently invalidates the perturbative expansion
for $\tau$ away from the initial time $\tau_0$. 

The key idea in the RG method, however, lies in the fact
 that we can utilize these apparently problematic
secular terms to obtain
an asymptotic solution valid in a global domain \cite{env001,env006}.
Indeed, one  can make the following
geometrical interpretation of the perturbative solution constructed around
arbitrary initial time $\tau_0$:
That is,
we have constructed a family of curves
$\tilde{f}_p(\tau \,,\, \sigma \,;\, \tau_0)$
parameterized with $\tau_0$,
which curves are all supposed to be on the exact solution
$f_p(\sigma;\, \tau)$ at $\tau = \tau_0$ up to $O(\varepsilon^3)$,
although they are admittedly only valid  for $\tau$ near $\tau_0$ locally.
Then,
the \textit{envelope curve} of the family of curves
will give a global solution in our asymptotic situation,
which is shown indeed to be the case \cite{env001,env006}.
According to the classical theory of envelopes,
the envelope that is in contact with any curve in the family
at $\tau = \tau_0$ is obtained \cite{env001} by
\begin{eqnarray}
  \label{eq:1-058}
  \frac{d}{d\tau_0}
  \tilde{f}_p(\tau , \sigma ; \tau_0) \Bigg|_{\tau_0 = \tau} = 0.
\end{eqnarray}
The derivative with respect to $\tau_0$ hits on
the hydrodynamic variables,
and hence we have 
the evolution equation of them, which is identified with the hydrodynamic 
equation \cite{env007,env009}.
We also note that the invariant manifold \cite{inv-manifold} 
corresponding to the hydrodynamics in the functional space of the distribution function
is explicitly obtained as 
an envelope function \cite{env009,Tsumura:2011cj}:
$f_{\mathrm{E}p}(\tau,\sigma) = \tilde{f}_p(\tau,\,\sigma \,;\, \tau_0 = \tau)$,
the explicit form of which is referred to Ref.'s \cite{env009,Tsumura:2011cj}.
We note that this solution is valid in a global domain 
of time in the asymptotic region \cite{Tsumura:2011cj}.
 
Putting back $\varepsilon$ to $1$,
Eq.(\ref{eq:1-058}) is reduced to the following form in this approximation,
\begin{eqnarray}
  \label{eq:1-061}
  \sum_{p}  \frac{1}{p^0}  \varphi_{0p}^\alpha 
  \Bigg[ (p \cdot u)  \frac{\partial}{\partial \tau}
  + p \cdot \nabla \Bigg]
  ( f^{\mathrm{eq}}_p +  \delta f^{(1)}_p ) = 0.\nonumber\\
\end{eqnarray}
where $\delta f^{(1)}_p$ denotes
the first-order correction to the distribution function
\begin{eqnarray}
  \label{eq:phi-bar}
  \delta f^{(1)}_p \equiv -\big[ A^{-1}  \bar{Q}_0  F \big]_p.
\end{eqnarray}
If one uses the identity
$(p \cdot u) \, \partial/\partial \tau
+ p \cdot \nabla = p^\mu\,\partial_\mu$,
Eq.(\ref{eq:1-061}) is found to have the following form
\begin{eqnarray}
  \label{eq:1-064}
  \partial_\mu T^{\mu\nu} = 0,\quad
  \partial_\mu N^{\mu} = 0,
\end{eqnarray}
with
$T^{\mu\nu} =T^{(0)\mu\nu}+\delta \, T^{\mu\nu}$ and
$N^{\mu}=N^{(0)\mu}+\delta\, N^{\mu}$.
Here, the zero-th order terms read 
  $T^{(0)\mu\nu}
  \equiv \sum_{p} \frac{1}{p^0}  p^\mu p^{\nu} f^{\mathrm{eq}}_p
  = e\,u^{\mu}\,u^{\nu}-p\,\Delta^{\mu\nu}$
and
$ N^{(0)\mu} \equiv \sum_{p}  \frac{1}{p^0}  p^\mu f^{\mathrm{eq}}_p = n\,u^{\mu}$,
with $e$, $p$, and $n$ being
the internal energy, pressure, and particle-number density
for the relativistic ideal gas, respectively,
while the dissipative parts are given by
\begin{eqnarray}
  \label{eq:1-065}
  \delta T^{\mu\nu}
  &\equiv& \sum_{p}  \frac{1}{p^0}  p^\mu p^{\nu}\delta f^{(1)}_p , \\
  \label{eq:delta-N}   
  \delta N^\mu
  &\equiv& \sum_{p}  \frac{1}{p^0}  p^\mu \delta f^{(1)}_p.
\end{eqnarray}

Note that the dissipative terms are 
due to the deviation $\delta f^{(1)}_p$ of the distribution 
function from the local one.
It is well known that
the local distribution function only gives
the (relativistic) Euler equation without dissipation.

\subsection{
   Uniqueness of Landau-Lifshitz frame
}

In this subsection,
we present  the dissipative parts
$\delta T^{\mu\nu}$ and $\delta N^{\mu}$, explicitly,
and discuss their properties.
An explicit evaluation of Eq.'s (\ref{eq:1-065}) and (\ref{eq:delta-N})
together with (\ref{eq:phi-bar})
gives \cite{env009,Tsumura:2011cj}
\begin{eqnarray}
  \label{eq:2-068}
  \delta  T^{\mu\nu} &=&  \zeta \,\Delta^{\mu\nu}\, \nabla\cdot u
  + 2 \, \eta \, \Delta^{\mu\nu\rho\sigma} \, \nabla_\rho u_\sigma,\\
  \label{eq:2-069}
  \delta N^\mu &=& \lambda \, \frac{1}{\hat{h}^2} \, \nabla^\mu\frac{\mu}{T},
\end{eqnarray}
respectively,
with 
$\Delta^{\mu\nu\rho\sigma} \equiv 
1/2 \cdot (\Delta^{\mu\rho}\Delta^{\nu\sigma} + \Delta^{\mu\sigma}\Delta^{\nu\rho}
- 2/3 \cdot \Delta^{\mu\nu}\Delta^{\rho\sigma})$.
Here,
$\hat{h}$ denotes
the reduced enthalpy per particle.
The bulk and shear viscosities 
and the thermal conductivity are denoted by $\zeta$, $\eta$ and $\lambda$, respectively. 
One readily finds that these formulas completely 
agree with those proposed by Landau and Lifshitz \cite{hen002}.
Indeed,
the respective dissipative parts
$\delta T^{\mu\nu}$ and $\delta N^{\mu}$
in Eq.'s (\ref{eq:2-068}) and (\ref{eq:2-069}) meet
 Landau and Lifshitz's constraints
\begin{eqnarray}
  \label{eq:delta-e}
   \delta e &\equiv& u_\mu  \,\delta T^{\mu\nu} \,u_\nu = 0, \\
  \label{eq:delta-Q}
  Q_\mu &\equiv& \Delta_{\mu\nu} \,\delta T^{\nu\rho} \,u_\rho = 0, \\
 \label{eq:delta-n}
  \delta n &\equiv& u_\mu \, \delta N^\mu = 0,
 \end{eqnarray}
which are imposed in a heuristic way in the phenomenological derivation \cite{hen002}.
Thus, we have found that
the frame on which the fluid velocity is 
defined necessarily becomes 
the Landau-Lifshitz (energy) frame,
 if the hydrodynamics is to be consistent with the underlying
RBE.

Next,
let us examine the underlying meaning of 
Eq.'s (\ref{eq:delta-e}) to (\ref{eq:delta-n})
in terms of the distribution function.
As was mentioned above,  these equations are usually just imposed \cite{mic001} 
to the higher-order terms of the distribution function  as 
the matching conditions without any foundation
 to select the hydrodynamic equation in the energy frame. 
We shall clarify that these conditions are 
equivalent to the orthogonality condition for the excited modes expressed 
in terms of the inner product
\cite{env009,Tsumura:2011cj} and hence an inevitable {\em consequence}
for the relativistic hydrodynamics in our analysis which is free from any ansatz.
 
We first note that Eq.'s (\ref{eq:1-065}) and (\ref{eq:delta-N}) can be rewritten as
\begin{eqnarray}
  \label{eq:1-065-2}
  \delta T^{\mu\nu}
  &=& \sum_{p}  \frac{1}{p^0}  p^\mu p^{\nu}f^{\mathrm{eq}}_p\bar{\phi}_p , \\
  \label{eq:delta-N-2}   
  \delta N^\mu
  &=& \sum_{p}  \frac{1}{p^0}  p^\mu f^{\mathrm{eq}}_p\bar{\phi}_p,
\end{eqnarray}
with
\begin{eqnarray}
  \bar{\phi}_p = (f^{\mathrm{eq}}_p)^{-1}\,\delta f^{(1)}_p = -\big[L^{-1}Q_0(f^{\mathrm{eq}})^{-1}F\big]_p,
\end{eqnarray}
which belongs to the Q$_0$ space and thus orthogonal to the zero modes,
\begin{eqnarray}
  \label{eq:4-005}
  \langle \, \varphi^\alpha_0 \,,\, \bar{\phi} \, \rangle = 0
  \,\,\,\mathrm{for}\,\,\,\alpha = 0,\,1,\,2,\,3,\,4.
\end{eqnarray}
Recalling the definition Eq. (\ref{eq:1-034}) of the inner product, 
we see that Eq. (\ref{eq:4-005})  with  $\alpha =\mu=0,\,1,\,2,\,3$ is reduced to
\begin{eqnarray}
  \label{eq:4-006}
  0&=& \sum_p  \frac{1}{p^0} (p \cdot u)  f^\mathrm{eq}_p 
  p^{\mu}\bar{\phi}_p 
  = u_{\nu}\sum_p  \frac{1}{p^0} p^{\nu}p^{\mu} f^\mathrm{eq}_p \bar{\phi}_p\nonumber\\
  &=& u_{\nu}\,\delta T^{\mu\nu},
\end{eqnarray}
which readily leads to Eq.'s(\ref{eq:delta-e}) and (\ref{eq:delta-Q}).
Quite similarly, Eq. (\ref{eq:4-005})  with  $\alpha=4$ is reduced to 
\begin{eqnarray}
  \label{eq:4-006-2}
  0 &=& \sum_p  \frac{1}{p^0} (p \cdot u)  f^\mathrm{eq}_p 
  \bar{\phi}_p 
  = u_{\nu}\sum_p  \frac{1}{p^0} p^{\nu} f^\mathrm{eq}_p \bar{\phi}_p\nonumber\\
  &=& u_{\nu}\,\delta N^{\nu},
\end{eqnarray}
which is nothing but Eq.(\ref{eq:delta-n}).

We emphasize again that 
the matching conditions for the energy frame
are  not imposed but
 uniquely obtained in our derivation from the underlying kinetic equation,
without any  assumptions nor ansatz.
This facts may mean that the relativistic hydrodynamic equation
for a viscous fluid must be defined in the energy frame, at least if it is
consistent with the underlying kinetic equation. 

\setcounter{equation}{0}
\section{
   Summary and Discussions
}
\label{sec:005}
In this paper,
we have shown that the renormalization-group (RG) derivation
of the relativistic dissipative hydrodynamic equation
as the infrared dynamics of 
the underlying relativistic Boltzmann equation (RBE) 
uniquely leads to the one in the energy frame proposed by Landau and Lifshitz,
provided that the macroscopic-frame vector, which
covariantly defines the local rest frame of the fluid velocity,
is independent of the momenta of constituent particles of
the system, as it should.

In relation to other methods of the derivation of hydrodynamic
equations based on the RBE,
we note that
we have not assumed any matching conditions 
\cite{mic001} but
uniquely got them in the energy frame 
from the underlying kinetic equation,
and hence
given the foundation for the 
matching conditions.
Since any matching conditions are not imposed,
secular terms due to the zero modes of the linearized collision operator 
appear inevitably in 
higher-order terms in our approach, but 
they are resummed 
away by the RG/envelope equation to give an asymptotic solution valid in a
global domain.

We thus argue that
the relativistic hydrodynamic equation
for  viscous fluids must be defined in the energy frame,
if it is consistent with the underlying kinetic equation. 
Although the RBE which we have adopted as the kinetic equation
is admittedly suitable only for a dilute gas, it is 
expected that the derived hydrodynamic equation itself and hence the uniqueness of the 
energy frame can be valid even for  
dense systems ;
this is found plausible if one recalls the universal nature  of
(non-relativistic) Navier-Stokes equation beyond dilute systems,
although it can be also derived  \cite{env007,chapman} from the (non-relativistic)
Boltzmann equation.

Although  
the present work is  confined  to the case of the so-called
first-order equation for a system composed of a single component,
the uniqueness of the energy frame for the relativistic hydrodynamics
may keep valid for the multi-component systems \cite{mic001}
and the case of the so-called second-order 
causal relativistic hydrodynamics,
\textit{i.e.},  the extended thermodynamics \cite{extended}.
In fact, we can show  that 
the energy frame is the most natural frame for multi-component systems \cite{multi}
and that the RG derivation \cite {next002}
of the mesoscopic dynamics \cite{meso} of the RBE
 naturally leads to the extended thermodynamics in the energy frame. 
We can thus assert more firmly  that
the relativistic hydrodynamic equations for viscous fluids
must be defined in the energy frame,  if it is consistent with the underlying
relativistic kinetic equation at all.


\begin{acknowledgments}
We are grateful to K. Ohnishi for his collaboration in the previous work, on which
the present work is based.
We thank  T. Hirano, A. Monnai
and Y. Hidaka, for
their inquiry on our previous work, in particular on  possible ambiguities
in the definition of the particle frame. We acknowledge that Y. Minami and Y. Hidaka
had a conjecture that the relativistic
hydrodynamic equation  might possibly be defined only in the energy frame, at least 
in the linear regime,  if
it should satisfy some basic thermodynamic property,
 which partly motivated the present work. 
T.K. thanks Y. Hidaka for the conversation on the latest status  of their work.
T.K. was partially supported by a
Grant-in-Aid for Scientific Research from the Ministry of Education,
Culture, Sports, Science and Technology (MEXT) of Japan
(Nos. 20540265 and 23340067),
by the Yukawa International Program for Quark-Hadron Sciences, and by a
Grant-in-Aid for the global COE program
``The Next Generation of Physics, Spun from Universality and Emergence'' from MEXT.
\end{acknowledgments}

\end{document}